\documentclass[12pt]{article}
\usepackage[utf8]{inputenc}
\usepackage{filecontents}
\usepackage{graphicx}
\usepackage{textcomp}
\usepackage{amsmath}
\usepackage{units}

\usepackage[style=nature]{biblatex}
\AtEveryBibitem{\clearfield{issn}}
\AtEveryCitekey{\clearfield{issn}}

\title{Few-nm tracking of magnetic vortex orbits and their decay with ultrafast Lorentz microscopy}

\author
{Marcel Möller$^{1}$, John H. Gaida$^{1}$, \\
Sascha Schäfer$^{1}$ and Claus Ropers$^{1,\ast}$\\
\\
\normalsize{$^{1}$4\textsuperscript{th} Physical Institute, University of Göttingen, 37077 Göttingen, Germany.}\\
\normalsize{$^\ast$To whom correspondence should be addressed; E-mail:  claus.ropers@uni-goettingen.de.}
}

\date{}

\begin{document}

\maketitle

\begin{abstract}
Transmission electron microscopy is one of the most powerful techniques to characterize nanoscale magnetic structures. In light of the importance of fast control schemes of magnetic states, time-resolved microscopy techniques are highly sought after in fundamental and applied research. Here, we implement time-resolved Lorentz imaging in combination with synchronous radio-frequency excitation using an ultrafast transmission electron microscope. As a model system, we examine the current-driven gyration of a vortex core in a 2\,µm-sized magnetic nanoisland. We record the trajectory of the vortex core for continuous-wave excitation, achieving a localization precision of $\pm2$\,nm with few-minute integration times. Furthermore, by tracking the core position after rapidly switching off the current, we find a temporal hardening of the free oscillation frequency and an increasing orbital decay rate attributed to local disorder in the vortex potential.
%In the future, time-resolved Lorentz microscopy will enable in-depth studies of current-driven magnetic phenomena in a versatile table-top transmission electron microscope. 
\end{abstract}

\section*{Introduction}
%Magnetic research has always topologically stabilized nanoscopic magnetic textures~\cite{Eggebrecht2017,Yu2018}

Magnetism gives rise to an incredibly rich set of nanoscale phenomena, including topological textures such as skyrmions~\cite{Skyrme1962,Muhlbauer2009,Yu2010} and vortices~\cite{Hubert1998}. Dynamical control of spin structures down to sub-picosecond timescales is facilitated by a multitude of interactions involving spin-torques~\cite{Ralph2008,Liu2011,Emori2013} or optical excitations~\cite{Beaurepaire1996,Stanciu2007,Kampfrath2011,Eggebrecht2017}. These features have shown immediate relevance in novel applications, exemplified in (skyrmion) racetrack memory~\cite{Parkin2008,Fert2013}, magnetic random access memory~\cite{Dave2005}, and vortex oscillators as radio-frequency  sources~\cite{Mancoff2005,Dussaux2010} or for neuromorphic computing~\cite{Macia2011,Torrejon2017}. 
Observing these magnetic phenomena on their intrinsic length and time scales requires experimental tools capable of simultaneous nanometer spatial and nanosecond to femtosecond temporal resolution.
Whereas photoelectron~\cite{Krasyuk2003} and magneto-optical schemes using pulsed radiation sources from the visible~\cite{Hiebert1997} to the x-ray~\cite{Fischer2003,VonKorffSchmising2014,Kfir2017} regime inherently offer high temporal resolution, the particular advantages of electron beam techniques have yet to be fully exploited in the ultrafast domain. 

Transmission electron microscopy (TEM) facilitates quantitative magnetic imaging with very high spatial resolution approaching the atomic level using aberration-corrected instruments~\cite{McVitie2015,Shibata2019} and holographic approaches~\cite{Lichte2008,Schofield2008,Midgley2009,Wolf2015}, while also allowing for structural or chemical analysis.
Importantly, the versatile  \textit{in-situ} environment of TEM enables the observation of nanoscale magnetic changes with optical~\cite{Eggebrecht2017,Fu2018,Berruto2018} or electrical stimuli~\cite{Pollard2012,Goncalves2017,Vogel2018}. Time-resolved transmission electron microscopy~\cite{Doomer2003,Zewail2010,Browning2012} has in some instances been used to address magnetization dynamics~\cite{Bostanjoglo1980,Park2010,Schliep2017,Fromter2017}.
The recent advance of highly coherent photoelectron sources~\cite{Feist2017,Houdellier2018} promises substantially enhanced contrast and resolution in ultrafast magnetic imaging~\cite{DaSilva2017}, but has, to date, not been combined with synchronized electrical stimuli.

In this work, we address this problem and realize ultrafast Lorentz microscopy with \textit{in-situ} radio-frequency (RF) current excitation. The technique is made possible by using a transmission electron microscope with a pulsed photoelectron source. Synchronizing the electron pulses at the sample with the radio-frequency current, images of the instantaneous state of the system for a given RF phase can be acquired. The controlled variation of the phase allows for recording high-speed movies with a temporal resolution independent of the readout speed of the detector, but given by the duration of the electron pulses and the quality of the electrical synchronization.

\section*{Results}

\subsection*{Time-resolved Lorentz Microscopy}

\begin{figure}[hbt!]
    \centering
    \includegraphics[width=89mm]{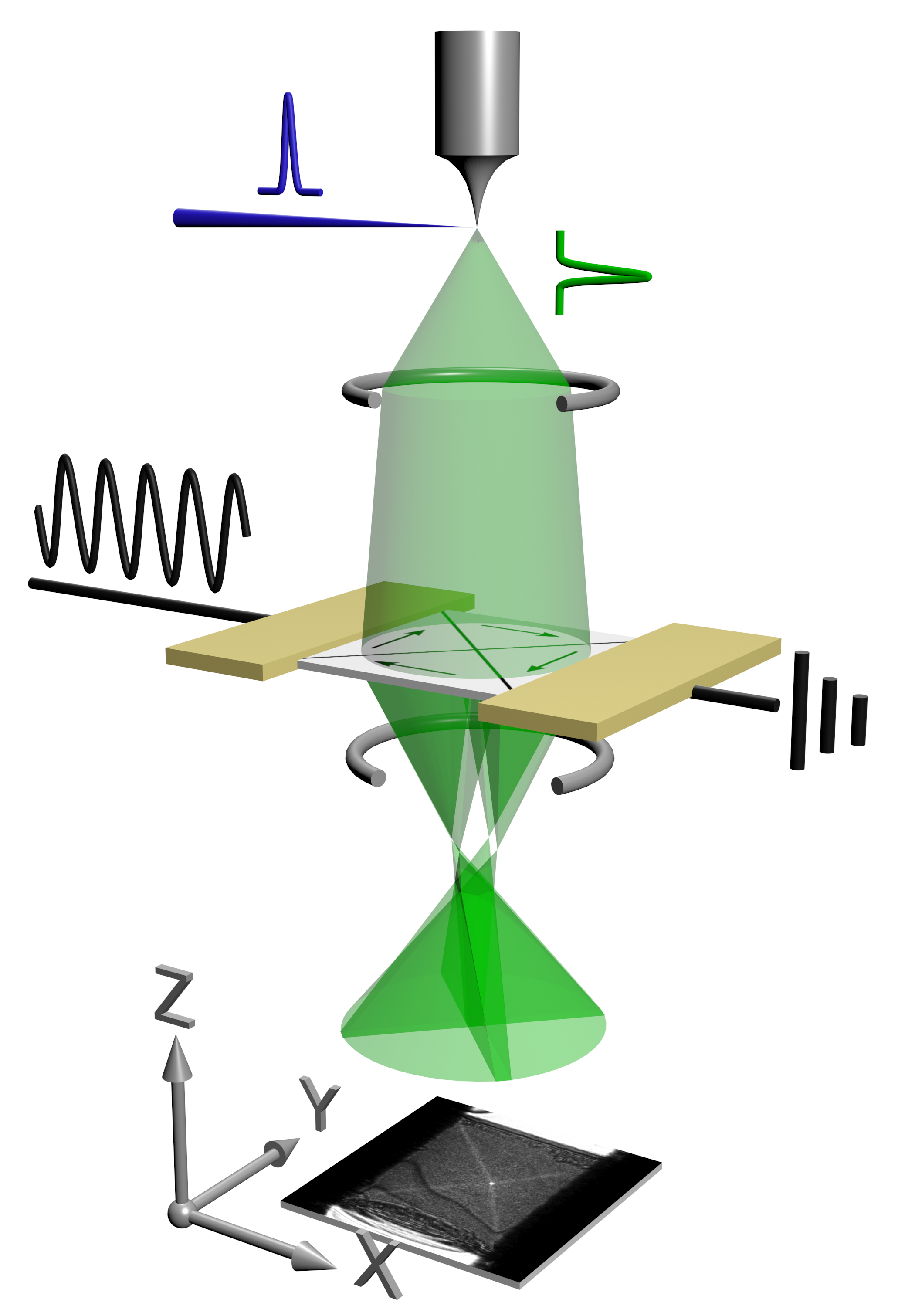}
    \caption{
    Schematic representation of time-resolved Lorentz microscopy combined with synchronous radio-frequency sample excitation: 
    An ultrashort electron pulse (green) is generated via a linear photoemission process from an optical laser pulse (blue) inside a TEM electron gun.
    A magnetic sample (light grey) is stroboscopically illuminated with a near-parallel electron beam and imaged onto a TEM camera.
    Employing a small defoucs in the imaging conditions gives rise to magnetic contrast (see text; exemplary image shown in the bottom).
    Dynamics in the sample are excited \textit{in-situ} with radio-frequency currents (black) phase-locked to multiples of the laser repetition rate, creating the appearance of a static image.
    A controlled change of the RF phase between images allows the entire dynamic process to be mapped in time.
    }
    \label{fig:my_label}
\end{figure}

The experiments were conducted at the Göttingen Ultrafast Transmission Electron Microscope (UTEM), which features ultrashort electron pulses of high beam quality from a laser-triggered field emitter~\cite{Feist2017}.
Employing a linear photoemission process, the UTEM allows for a flexible variation of the electron pulse parameters in terms of repetition rate and pulse duration~\cite{Feist2017,Bach2019}. In the present study, we employ electron pulses at a repetition rate of 500 kHz, driven by optical pulses from an amplified Ti:Sapphire laser with a central wavelength of 400\,nm and a pulse duration of 2.2\,ps. %The RF current is synchronized with the electron pulses using a fast photodiode signal from the laser oscillator.

We demonstrate the capabilities of our instrument by studying the model system of a spin-transfer torque driven vortex oscillator~\cite{Choe2004,Bolte2008,Bukin2016,Bisig2016,Fromter2017}.
The sample is a 2.1$\times$2.1\,µm$^2$ large and 20\,nm thick polycrystalline permalloy (Ni$_{81}$Fe$_{19}$) square.
Its magnetic configuration consists of four domains forming a flux closure (Landau) state with an in-plane magnetization pointing along the edges of the nanoisland and a perpendicular magnetized core with a diameter on the order of 10\,nm~\cite{Hubert1998,Okuno2000}. 
The chirality $c$ specifies the counter-clockwise ($c = +1$) or clockwise ($c = -1$) in-plane curling direction, whereas the polarity $p$ indicates if the core magnetization is parallel ($p = 1$) or antiparallel ($p = -1$) to the $z$-axis of the coordinate system defined in Fig. 1.
%The magnetic domain configuration of the nanoisland forms a flux-closure (Landau) state curling around the center of the square, thereby minimizing magnetic stray fields~\cite{Hubert1998,Okuno2000}. 
%The curling direction, called the chirality $c$, can either be c.
%In its center, the vortex core, the magnetization is directed out of the sample plane. The polarity $p$ of the vortex core indicates if the magnetization points parallel ($p = 1$) or antiparallel ($p = -1$) to the $z$-axis of the coordinate system defined in Fig. 1. 

Two edges of the structure are electrically contacted with 100\,nm thick gold electrodes (see Fig.~1) which terminate in wire-bonding pads. The sample is installed in a custom-made TEM holder that allows for \textit{in-situ} RF-current excitation up to the GHz-regime~\cite{Pollard2012,Goncalves2017,Vogel2018}.
RF-currents are generated with an arbitrary waveform generator (AWG) synchronized to the photodiode signal from the laser oscillator. This allows us to create custom waveforms, such as continuous or few-cycle sine waves, with a fixed, software-programmable timing between the probing electron pulse and the phase of the excitation.
Driven by alternating currents, the vortex structure exhibits a resonance behaviour, in which the core traces out an ellipsoidal trajectory~\cite{Kruger2007,Kruger2010,Pollard2012,Bisig2016}.
The trajectory of the vortex core under the simultaneous influence of spin-transfer torques and Oersted fields is predicted by a harmonic oscillator model~\cite{Kruger2007,Kruger2010}.

To image the magnetic configuration of our sample, we employ Fresnel-mode Lorentz microscopy~\cite{DeGraef2001a,Zweck2016}, a form of defocus phase contrast (see Fig.~1). In a classical picture, magnetic contrast arises from the Lorentz force:
When the parallel beam of electrons passes through the magnetic field of the sample, electrons experience a momentum change perpendicular to their propagation direction and proportional to the local magnetic field.
By slightly defocusing the imaging system of the microscope, this transverse momentum gives rise to a contrast which scales with the magnitude of the gradient of the sample's in-plane magnetization~\cite{DeGraef2001a,Zweck2016}. 

A Lorentz micrograph of the sample acquired with a continuous electron beam is depicted in the bottom of  Fig.~1 (see also Methods Fig.~4\,a).
The X-shaped lines indicate the position of the domain walls. 
A peak in the contrast arises at the intersection of the domain walls, and marks the position of the vortex core. The core itself does not contribute to the magnetic contrast, as its magnetization points along the beam propagation direction.
The dark lines in the lower half of the image are Bragg lines of the single-crystal silicon substrate, which locally scatter intensity out of the forward beam.

\subsection*{Continuous Excitation}

\begin{figure*}[hbt!]
    \centering
    \includegraphics{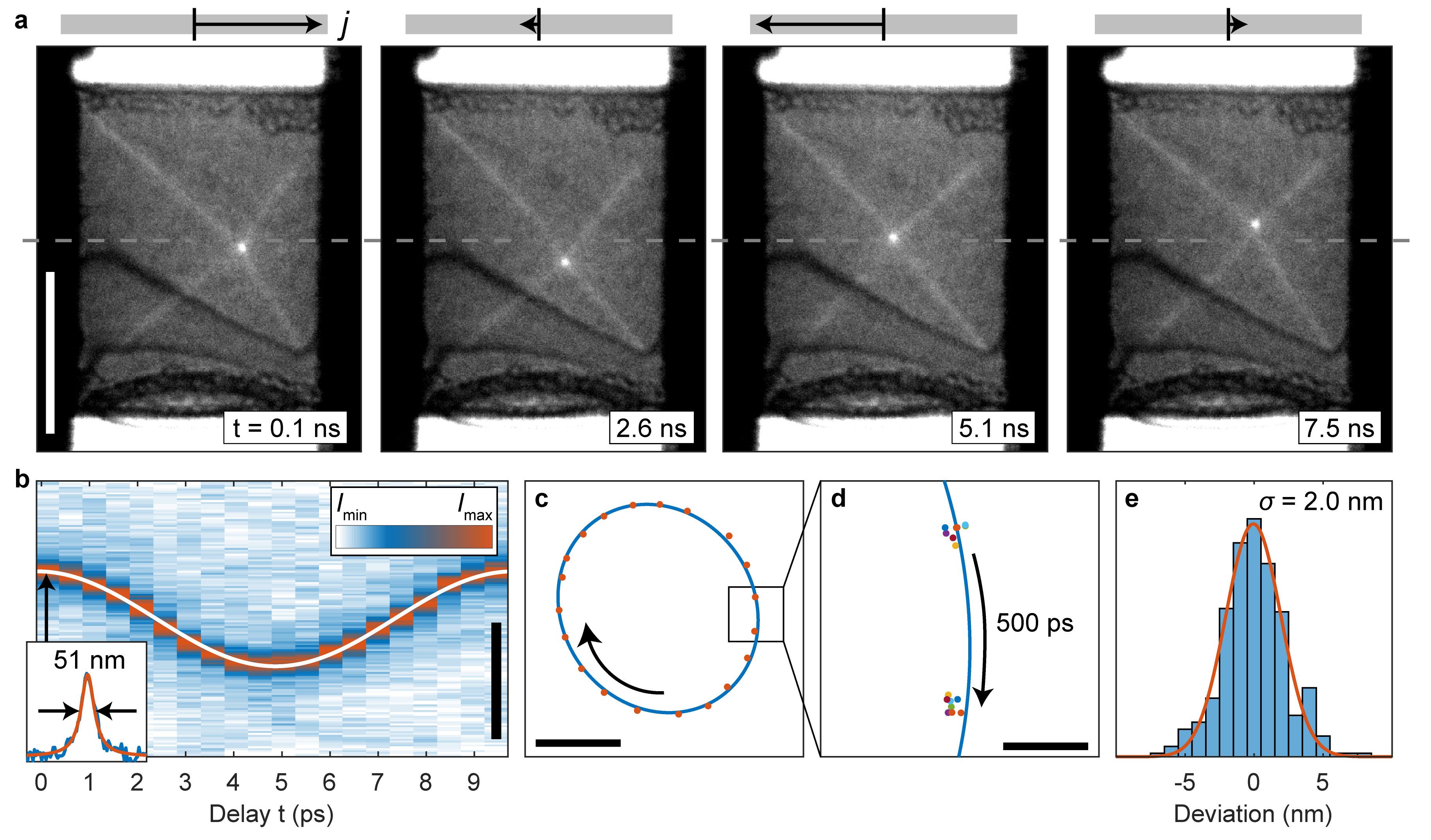}
    \caption{
    Continuous-wave excitation at a frequency of $f_{ex} = 101.5$\,MHz and a current density of $j = 1.7 \cdot 10 ^{11}$ A/m$^2$.
    (a)~Four time-resolved Lorentz micrographs acquired in time steps of 2.5\,ns at a defocus of 520\,µm (scale bar: 1\,µm).
    (b)~Line profiles at the position of the vortex core along the $x$-direction, closely reassembling a sinusoidal displacement. White: Best fit of a harmonic oscillator model (scale bar: 300\,nm).
    (b, Inset)~Blue: Example of a single line profile at $t = 0.1\,$ps. Red: Fitted Lorentzian line shape.
    (c)~Red: Mean vortex core position determined from set of 8 identical Lorentz micrographs at each delay $t$. Blue: Best fit of the harmonic oscillator model (scale bar: 100\,nm).
    (d)~Example of the spatial spread of the tracked vortex position (scale bar: 20\,nm).
    (e)~Distribution of the spatial deviation from the mean for all position measurements in $x$- and $y$-direction.
    }
\end{figure*}

In a first experiment, we excite the vortex with a continuous sinusoidal current at a frequency of $f_{ex} = 101.5\,$MHz.
This frequency maximizes the orbit diameter of the gyration, as determined from a preliminary measurement using a continuous electron beam (see Methods). To resolve a whole period, we acquire images for 21 delays $t$ at intervals of 500\,ps. 
At every delay $t$, we record 32 time-resolved micrographs, each with an integration time of one minute. 
The delay times given are quantitative with respect to the maximum of the driving field, enabled by \textit{in-situ} electron beam deflection near the contacts (see Methods)~\cite{Goncalves2017}.
After applying a drift-correction, we sum over identical micrographs and obtain a movie of the vortex dynamics, which can be found in the supplementary material (Supplementary movie 1).

Between the four exemplary frames in Fig.~2\,a, acquired at time steps of 2.5\,ns, the vortex core is clearly displaced and performs a clockwise rotation, implying a polarity of $p = -1$~\cite{Choe2004,Kruger2007}.
The chirality of the vortex is $c = -1$, determined from the contrast of the domain walls (bright or dark; here: bright) and the choice of imaging conditions (over- or underfocus; here: overfocus)~\cite{Schneider2000}.
Line profiles along the $x$-direction at the position of the vortex core are displayed in Fig. 2\,b, following a near-sinusoidal trajectory.
The inset in Fig. 2\,b depicts a single line profile (blue line) with a Lorentzian line-shape (red line) fitted to the data.
The average full-width-half-max (FWHM) of all line profiles in the $x$- and $y$-directions is $51 \pm 1\,$nm, which represents an upper limit for the magnetic point resolution and constitutes the best spatial resolution of stroboscopic Lorentz-microscopy to date.

Due to the high signal-to-noise ratio, the position of the vortex core can be tracked with a precision far below the point resolution, as shown in the following.
%Figure 2c displays the position of the vortex core during one oscillation period.
From the 32 images for every delay $t$, we form a set of eight identical images, each corresponding to an integration time of 4 minutes.
In a single image, the position of the vortex core is determined by the center of mass of the peak.
To visualize the position of the vortex core during one oscillation period, we average over the eight individual measurements (Fig.~2\,c).
An example of their distribution is shown in Fig.~2\,d.
%The red dots in Fig. 2c correspond to the mean value of these eight individual measurements, whereas Fig. 2d shows an example of their distribution.
One can clearly see that the points accumulate in an area with a diameter of less than 10\,nm.
This is significantly smaller than the FWHM of the peak in the image contrast.
The deviations from the mean of all position measurements are combined in Fig.~2\,e.
The histogram follows a Gaussian distribution with a standard deviation of $\sigma = 2.0$\,nm, corresponding to the localization precision of $\pm2\,$nm in each spatial dimension. 

Knowledge of the absolute timing of the vortex gyration with respect to the driving current allows us to directly fit our results to the oscillator model of Ref.~\cite{Kruger2010} (see Methods).
Although the fitted trajectory agrees well with the data (cf. white and blue lines in Figs.~2\,b,\,c), the obtained resonance parameters of the free oscillation frequency $f_0= 97.9 \pm 2.1$\,MHz and the damping $\Gamma = 74 \pm 13$\,MHz indicate a limitation of the description.
Specifically, the damping appears unrealistically high, given the response curve in Fig.~4\,b and previous measurements \cite{Pollard2012}. This suggests that additional contributions in the restoring force or the damping may affect the vortex motion, arising, for example, from local disorder~\cite{Compton2006, Compton2010,Min2010,Min2011,VanDeWiele2012,Leliaert2014} or global anharmonicities in the vortex potential~\cite{Guslienko2010,Drews2012,Stevenson2013,Guslienko2014}. In order to obtain direct time-domain information on the damping and possible deviations from a harmonic confinement, we conduct measurements of the free-running relaxation of the vortex in the absence of a driving force.

\subsection*{Damped Motion}

\begin{figure}[hbtp!]
    \centering
    \includegraphics{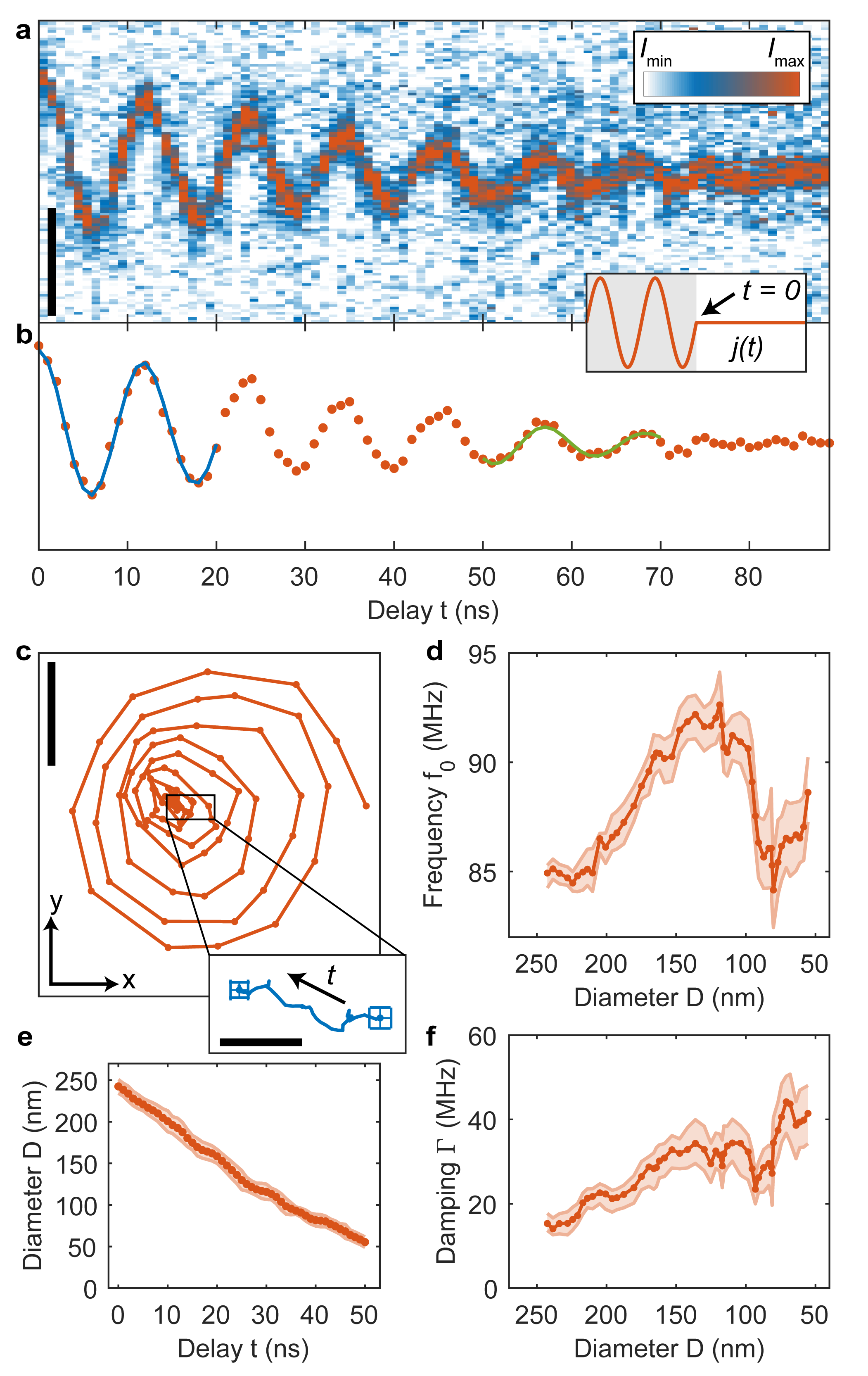}
    \caption{
    Damped vortex gyration after switching off the current ($f_{ex} = 101.5\,$MHz, $j = 2.9 \cdot 10 ^{11}$\,A/m$^2$).
    (a)~Profiles of the vortex core along the $x$-axis, illustrating the damped sinusoidal motion towards the equilibrium position (scale bar: 200\,nm).
    (b)~Red: Tracked $x$-position of the vortex core as function of $t$. Blue/green: First/last result of the moving-window fit to the vortex trajectory.
    (c)~Spatial vortex core trajectory during the damping. Spacing between the dots is 1\,ns (scale bar: 100\,nm).
    (c,\,Inset)~Center of the gyration moving over time as determined from the fit (scale bar: 20\,nm).
    (d,\,e,\,f)~Fit results of the windowed fit. Shaded regions are one-sigma confidence intervals:
    (d)~Free oscillation frequency $f_0$ as a function orbit diameter $D$.
    (e)~Mean orbit diameter $D$ over time $t$.
    (f)~Damping $\Gamma$ as a function of orbit diameter $D$.
}
\end{figure}

In the following measurements performed at the same sample, a current at an identical frequency of $f_{ex} = 101.5\,$MHz is applied for a duration of 1\,µs, and is switched off at a zero-crossing of the sine wave at $t=0$ (see inset of Fig.~3\,a). 
Again, the timing was directly determined via small-angle scattering in the vicinity of the permalloy square.

At delay steps of 1\,ns, we acquire Lorentz micrographs with an integration time of 3\,min. An assembled movie of the data is found in the supplementary material (Supplementary movie M2). The path of the core to its rest position is clearly visible both in the line profiles along the $x$-axis (Fig.~3\,a) and in the tracked position of the vortex core (Fig.~3\,b).
A two-dimensional representation of the complete trajectory is given in Fig.~3\,c.
As can be seen from the sense of rotation in Fig.~3\,c, the vortex polarity is $p = +1$ for this measurement. Thus, we chose a slightly higher current density of $j = 2.9 \cdot 10 ^{11}$\,A/m$^2$, which compensates for the smaller orbits in case of an opposite sign between chirality and polarity (cf. Eq.~2)~\cite{Kruger2010}.

The trajectory obtained can not be expressed in terms of a single exponentially decaying sinusoidal function, illustrating systematic deviations from the damped harmonic oscillator model. However, we can extract the momentary frequency and decay rate of the oscillation by fitting an exponentially decaying orbit in a moving window with a size of 20\,ns.
The blue and green curves in Fig.~3\,b display the fitted $x$-component for windows starting at $t_w= 0$\,ns and $t_w= 50$\,ns, respectively.
Good agreement was also obtained for all intermediate values.

The results of the fit are summarized in Figs.~3\,d-e. 
The temporal evolution of the orbit diameter $D$ exhibits an approximately linear rather than exponential decay (Fig.~3\,e), evidenced by the increase of the instantaneous damping $\Gamma$ (Fig.~3\,f).
The trajectory further experiences a temporary hardening of the momentary frequency $f_0$ (Fig.~3\,d) from 85 to about 92\,MHz at medium orbit diameters $D$.
Moreover, allowing for a variable gyration center, we find that the windowed fit reveals a continuous translation of the orbit towards the left (inset in Fig.~3\,c).

The transient acceleration of the gyratory motion implies the presence of disorder in the vortex potential, and is consistent with time-resolved optical studies~\cite{Compton2006,Compton2010}.
Similarly, the movement of the orbit center suggests a trapping of the vortex core in a pinning site and is further supported by the broadening of the profile in Fig.~3\,a for delays $t > 80$\,ns, likely caused by a stochastic motion at very small diameters. As mentioned above, such frequency shifts have been characterized by other approaches~\cite{Compton2006,Compton2010,Dussaux2010}. Yet, we have not encountered a real-space measurement of the type of damping observed here.
A micromagnetic theoretical study by Min \textit{et al.}~\cite{Min2011} points towards a possible microscopic origin of the enhanced dissipation. In that work, spatial disorder in the form of variations in the saturation magnetization was found to cause additional damping via deformations of the internal magnetic structure of the vortex.
While our observations, in agreement with the theoretical calculations, show an increase in $f_0$ and $\Gamma$, particular differences are apparent.
Specifically, the damping growth is much more pronounced in our case, such that the ratio $\Gamma/f_0$ increases over time, while it is found to decrease in Ref.~\cite{Min2011}. Therefore, our measurements suggest additional contributions to the disorder potential, such as magnetic anisotropy or non-magnetic voids~\cite{VanDeWiele2012}.
A detailed modeling may be guided by further correlative studies of the structural and chemical composition, facilitated by nanoscale diffraction and spectroscopy available in transmission electron microscopy.

In conclusion, we implemented ultrafast Lorentz microscopy with synchronous radio-frequency current excitation and demonstrated its high spatio-temporal resolution by mapping the time-resolved gyration of a magnetic vortex core with an precision of $\pm2.0$\,nm.
Future developments will include higher contrast and spatial resolution from enhanced beam coherence, increased sensitivity from larger duty-cycles and frequencies up to the terahertz range.
In our opinion, exposing spin textures, such as vortices and skyrmions, to electrical and electromagnetic stimuli and observing the nanoscopic response will make valuable contributions to fundamental research and device development in ultrafast magnetism.
Moreover, the approach can be readily extended to \textit{in-situ} phase contrast imaging in different areas, including electrical switching of multiferroic materials and other correlated systems and heterostructures.

\section*{Methods}

\subsection*{Instrumentation}
The data was acquired at a JEOL 2100F TEM, which is equipped with a Gatan Ultrascan US4000 camera and features an electron gun allowing for a photoemission of electron pulses. Details concerning the electron source can be found in Ref.~\cite{Feist2017}. The laser system in use is a Ti:Sa regenerative amplifier and seed laser manufactured by Coherent, which outputs 35\,fs pulses at a repetition rate of $f_{rep} = 500$\,kHz and a wavelength of 800\,nm.
The output of the amplifier is frequency doubled using a $\beta$-Barium borate crystal and subsequently stretched to 2.2\,ps in a 10\,cm SF6 bar before it is focused onto the electron source.

The amplifier system provides an electrical trigger signal which is synchronized to its optical output. This signal is fed into the trigger input of a Keysight 81160A arbitrary waveform generator, where it initiates the output of a sinusoidal wave with a fixed number of cycles $n_{Burst}$. For a quasi-continuous excitation of the sample, the burst number and the excitation frequency are set to the same multiple of the repetition rate of the laser amplifier, i.e. $f_{ex} = n_{Burst}\cdot f_{rep}$. For non-continuous excitations, there is no constraint on the waveform other than it being shorter than $1/f_{rep}$. 

In quasi-continuous operation special care has been taken to achieve a glitch- and gap-free output by monitoring the signal with a directional coupler (Mini-Circuits ZFDC-20-4L) and an oscilloscope (Tek DPO71604C with 16 GHz bandwidth). Additionally, we placed a 6\;dB attenuator after the generator output to suppress standing waves at the RF transmission line, originating from the unmatched sample and a parasitic capacitance of the generator output.

\subsection*{Sample System and Preparation}
\begin{figure}[hbt!]
    \centering
    \includegraphics{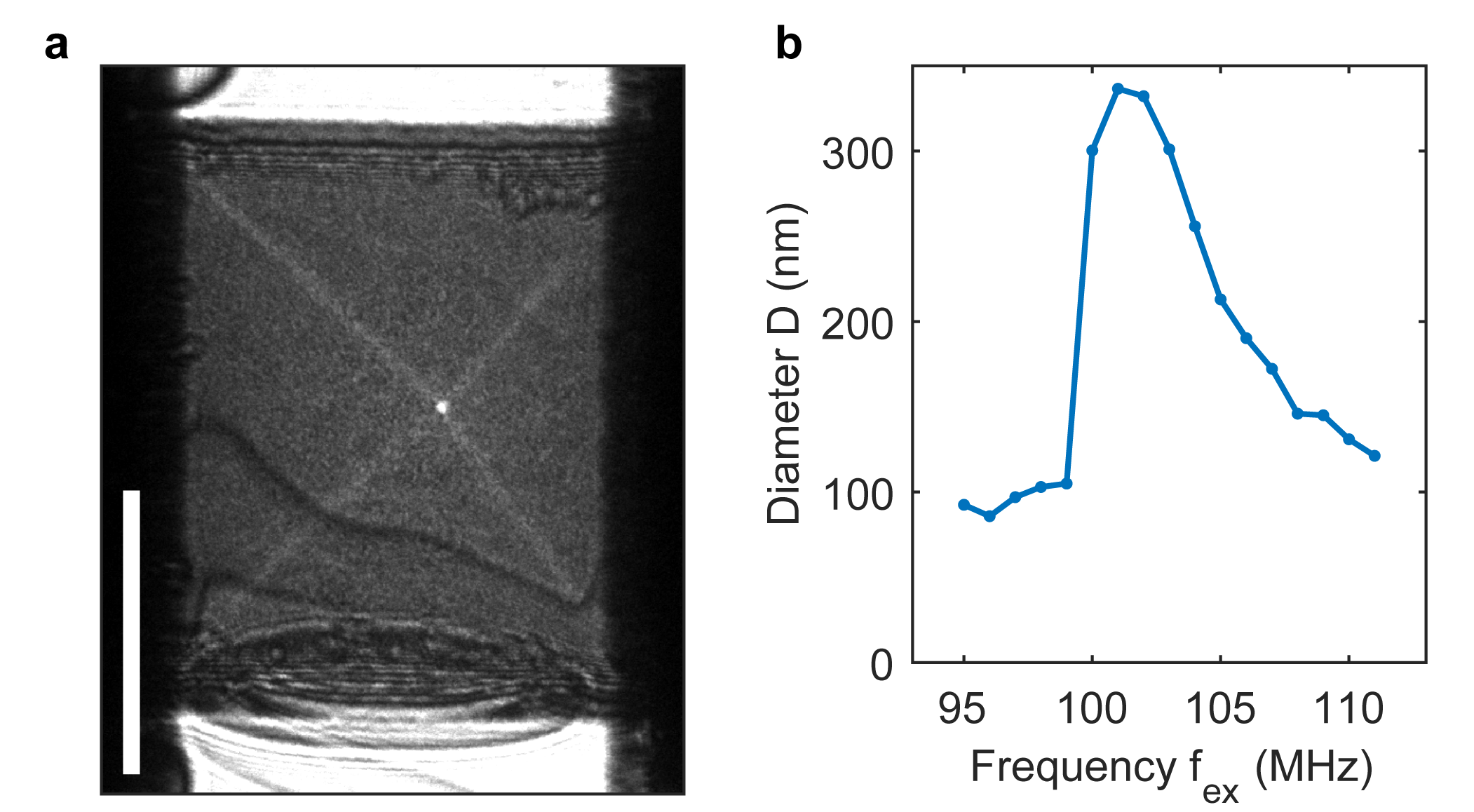}
    \caption{
    Preliminary characterization of the sample utilizing an continuous electron beam.
    (a) Lorentz-micrograph of the sample system acquired under continuous illumination.
    (b) Orbit-diameter of the vortex gyration as a function of the driving frequency~$f_{ex}$.
    }
\end{figure}

The sample consists of a 20\,nm thick permalloy (Ni$_{81}$Fe$_{19}$) square of $2.1 \times 2.1\,$\textmu m$^2$ size, which is electrically connected with two gold contacts (100\,nm thickness). The substrate is a single crystalline, 15\,nm thick silicon membrane with $100 \times 100\,$\textmu m$^2$ windows. 
The magnetic structure and the contacts were fabricated using electron beam lithography and thermal evaporation.
A Lorentz micrograph recorded with a continuous electron beam is depicted in Fig.~4\,a.
The resonance curve of the vortex oscillator (Fig. 4\,b) was determined from a Lorentz image series acquired prior to the time-resolved measurement using a continuous electron beam (cf. Ref.~\cite{Pollard2012}). 
A maximum response near $f_{ex} = 101.5\,$MHz is found.

\subsection*{Tracking the vortex core}
To track the position of the vortex core, we denoised the frames by applying a 3\,x\,3 pixel$^2$ (4.85\,nm/px) median filter and calculated the center of mass of the pixels of maximum intensity. These pixels were selected via thresholding. The threshold is chosen such that the contributing pixels correspond to a region of about 50\,nm in diameter.

\subsection*{Determination of time-zero}
We utilize small-angle electron scattering to determine the absolute timing of the current through the nanoisland by measuring the electrical field between the cold contacts~\cite{Goncalves2017}.
Therefore, we form a parallel electron beam close to an edge of an Au electrode in a sufficient distance to the permalloy. This ensures that the electrons are only deflected due to the electric field between the contacts.
In diffraction mode, this beam converges to a single point. Choosing a long camera length, we can image a deflection of this spot which is directly proportional to the electric field.

To determine $t=0$ of the first measurement (continuous sinusoidal excitation), we fit a cosine-function to the delay-dependent position of the electron spot.
For the second measurement of the damped motion, we probe delays before and after $t=0$, which allows us to directly find the last zero-crossing of the sinusoidal current.

\subsection*{Harmonic Oscillator Model}
%The harmonic oscillator model for the movement of a vortex core in a square magnetic thin film used here was published by Krüger et al. and is based on the Landau-Lifshitz-Gilbert (LLG) equation. 
Passing a current through a non-magnetic/magnetic (here: Au/Py) interface spin polarizes this current due to a difference in the scattering probability between the majority and minority electrons in the conduction band of the magnetic layer~\cite{Ralph2008}.
This spin-polarized current exerts a torque on the magnetic domains, altering their configuration and resulting in a displacement of the vortex core. Zhang and Li derived a general micro-magnetic model accounting for adiabatic and nonadiabatic contributions to this spin-transfer-torque (STT) by extending the the Landau-Lifshitz-Gilbert (LLG) equation~\cite{Zhang2004}.
Based on a generalized Thiele equation~\cite{Thiele1973,Thiaville2005}, Krüger \textit{et al.} published a harmonic oscillator model for the movement of a vortex core in a square magnetic thin film, which takes in-plane Oersted fields as well as spin-transfer torques into account. These Oersted fields originate in an inhomogeneous current density inside the magnetic conductor~\cite{Kruger2007,Bolte2008,Kruger2010}.
Krüger \textit{et al.} specify the resulting steady state trajectory under the influence of a driving current $ \vec j (t)  \propto \exp \left(i \Omega t\right)\vec e_x$ and field $ \vec H(t) \propto \exp \left( i \Omega t\right) \vec e_y$ in the supplementary material of Ref.~\cite{Kruger2010} as:

\begin{align}
    \begin{pmatrix}
    X\\
    Y
    \end{pmatrix}
    =
%    &A
%    \begin{pmatrix}
%    i \\
%    p
%    \end{pmatrix}
%    e^{-\Gamma t +i \omega t}
%    + B
%    \begin{pmatrix}
%    -i \\
%    p
%    \end{pmatrix}
%    e^{-\Gamma t -i \omega t}
%    \\
    &- \frac{e^{i\Omega t}}{\omega^2 + (i \Omega + \Gamma)}
    \begin{pmatrix}
        \Tilde{j} & \Tilde{H}cp + \left|\frac{D_0}{G_0}\right|p \xi \Tilde{j} \\
        - \Tilde{H} cp - \left|\frac{D_0}{G_0}\right|p\xi\Tilde{j} & \Tilde{j}
    \end{pmatrix}
    \begin{pmatrix}
        \frac{w^2}{w^2+ \Gamma^2}i\Omega \\
        \omega p + \frac{\omega\Gamma}{\omega^2 + \Gamma^2}i \Omega p
    \end{pmatrix}
    \label{eq:steady-state-trajectory}
\end{align}

with $\Tilde{H}$ and $\Tilde{j}$ the normalized magnitudes of the Oersted-field and the spin-polarized current, $\Omega = 2\pi f_{ex}$ the driving frequency, $\omega = 2\pi f_0$ the free oscillation frequency, $\Gamma$ the damping constant and $\left|D_0/G_0\right|$ a ratio related to exact magnetic distribution inside the nanoisland. The non-adiabicity parameter $\xi$ describes the ratio between non-adiabatic and adiabatic contributions in the LLG equation.

The scalar prefactor together with the vector define a trajectory in the form of an ellipse.
Due to its symmetry, the matrix in (\ref{eq:steady-state-trajectory}) only introduces a rotation and an isotropic dilation to this ellipse. Parameters such as the ellipticity and the phase of the gyration (i.e., the position of the core with respect to the major axis of the ellipse  at $t=0$) are independent of this matrix. 
Therefore, we can simplify Eq. (1) to 
\begin{align}
    \begin{pmatrix}
    X\\
    Y
    \end{pmatrix}
    =
    - \frac{e^{i\Omega t}}{\omega^2 + (i \Omega + \Gamma)}
    \;A\;\hat{R}(\theta)\cdot
    \begin{pmatrix}
        \frac{w^2}{w^2+ \Gamma^2}i\Omega \\
        \omega p + \frac{\omega\Gamma}{\omega^2 + \Gamma^2}i \Omega p
    \end{pmatrix}
    \label{eq:steady-state-trajectory-simplified}
\end{align}
with an isotropic dilation $A$ and a rotation of the ellipse by angle $\theta$ defined by the rotation matrix $\hat{R}(\theta)$.
From the vector in Eq. (2) it follows directly that the sense of the rotation depends only on the core polarization $p$, with $p=+1/-1$ resulting in a counter-clockwise/clockwise rotation \cite{Bolte2008}.

The trajectory in Eq.~(2) has five measurable parameters, which equals the number of degrees of freedom of an ellipse (length of the two semiaxes, tilt, frequency) plus an initial phase.
This allows us to directly fit the model to our measurement of the continuously driven oscillation by solving the overdetermined system of 42 equations ($X(\Delta t)$ and $Y(\Delta t)$ for every delay $\Delta t$).
As the driving frequency is known, we fix it to $f_{ex} = 101.5$\,MHz.
The best fit ($f_0= 97.9 \pm 2.1$\,MHz, $\Gamma = 74 \pm 13$\,MHz) is shown as solid line in Fig.~2\,b (white) and Figs.~2\,c,\,d (blue). The errors in the fitted values correspond to $1 \sigma$-confidence intervals determined by bootstrapping~\cite{Efron1992}.

%\begin{align}
%    \begin{pmatrix}
%    X(t)\\
%    Y(t)
%    \end{pmatrix}
%    =
%    \frac{A}{2}
%    \begin{pmatrix}
%    \sin (\varphi - \omega t) \\
%    p\cdot\cos(\varphi - \omega t)
%    \end{pmatrix}
%    e^{-\Gamma t}
%    + 
%    \begin{pmatrix}
%    X_0\\
%    Y_0
%    \end{pmatrix}\;,
%\end{align}

\section*{Acknowledgement}
We thank Johannes Riebold for the design and testing of the \textit{in-situ} TEM Holder.
Furthermore, we acknowledge useful discussions with Mathias Kläui, Henning Ulrichs and Michael Vogel and assistance from the Göttingen UTEM Team, especially Armin Feist, Nara Rubiano da Silva, Nora Bach, Thomas Danz and Karin Ahlborn.
This work was funded by the Deutsche Forschungsgemeinschaft (DFG) in the Collaborative Research Center ``Atomic Scale Control of Energy Conversion'' (DFG-SFB 1073, project A05) and via resources from the Gottfried Wilhelm Leibniz-Price.
We gratefully acknowledge support by the Lower Saxony Ministry of Science and Culture and funding of the instrumentation by the DFG and VolkswagenStiftung.

%%%%%%%%%%%%%%%%%%% FOR SUBMISSION %%%%%%%%%%%%%%%%%%%%%%%%
% Comment the two following commands
% Copy the results of the .bbl file  here
\printbibliography

\end{document}